\documentclass[aps,prb,twocolumn,showpacs]{revtex4-1}
\usepackage{graphicx}
\usepackage{graphics}
\usepackage{amsmath}
\usepackage{amsfonts}
\usepackage{amssymb}
\usepackage{epstopdf}
\usepackage{makeidx}
\usepackage{epsfig}
\usepackage{bm}
\usepackage{color}
\usepackage{xcolor}
\usepackage[unicode=true, bookmarks=false,breaklinks=false,pdfborder={0 0 1},backref=false,colorlinks=false]{hyperref}
\hypersetup{colorlinks=true,citecolor=blue,linkcolor=blue,filecolor=blue,urlcolor=blue}

\begin{document}

\newcommand{\be}   {\begin{equation}}
\newcommand{\ee}   {\end{equation}}
\newcommand{\ba}   {\begin{eqnarray}}
\newcommand{\ea}   {\end{eqnarray}}
\newcommand{\ve}   {\varepsilon}
\newcommand{\CAIO}[1]{\textcolor{red}{\fbox{Caio} {\sl#1}}}
\newcommand{\leandro}[1]{\textcolor[rgb]{1,0,1}{\fbox{Leandro} {#1}}}
\title{Random gauge field effects on the conductivity of\\ graphene sheets with disordered ripples}
\author{Rhonald Burgos}
\affiliation{Instituto de F\'{\i}sica,
  Universidade Federal Fluminense, 24210-346 Niter\'oi RJ, Brazil}
\author{Jesus Warnes}
\affiliation{Instituto de F\'{\i}sica,
  Universidade Federal Fluminense, 24210-346 Niter\'oi RJ, Brazil}
\author{Leandro R. F. Lima}
\affiliation{Instituto de F\'{\i}sica,
  Universidade Federal Fluminense, 24210-346 Niter\'oi RJ, Brazil}
\author{Caio Lewenkopf}
\affiliation{Instituto de F\'{\i}sica,
  Universidade Federal Fluminense, 24210-346 Niter\'oi RJ, Brazil}
 
\date{\today}

\begin{abstract}
We study the effect of disordered ripples on the conductivity of monolayer graphene flakes. 
We calculate the relaxation times and the Boltzmann conductivities associated with two mechanisms. 
First, we study the conductivity correction due to an external in-plane magnetic field $B_{\parallel}$. 
Due to the irregular local curvature found at graphene sheets deposited over a substrate, $B_{\parallel}$ 
can be mapped into an effective random magnetic field perpendicular to the graphene surface. 
Second, we study the electron momentum relaxation due to intrinsic pseudo magnetic fields originated
from deformations and strain.
We find that the competition between these mechanisms gives rise to a strong anisotropy in the 
conductivity tensor. This result provides a new strategy to quantitatively infer the strength of 
pseudo-magnetic fields in rippled graphene flakes.
\end{abstract}

\pacs{72.80.Vp,73.23.-b,72.10.-d}

\maketitle

\section{Introduction} 
\label{sec:purpose}

The electronic transport properties of bulk graphene are quite remarkable 
\cite{CastroNeto09,Mucciolo10,Peres2010,DasSarma11}. 
At room temperature graphene has mobilities \cite{Geim2007} as high as 
$\mu\approx15,000$ cm$^2$V$^{-1}$s$^{-1}$, that are significantly larger than any 
other semiconductor. At low temperature, however, the typical mobilities increase
only up to $\mu\approx200,000$ cm$^2$V$^{-1}$s$^{-1}$, which is a disappointing 
figure as compared with high quality GaAs heterostructures. These observations 
triggered an intense theoretical and experimental activity to understand the disorder 
mechanisms that limit the mobility in graphene (see, for instance, 
Refs.~\onlinecite{CastroNeto09,Mucciolo10,Peres2010,DasSarma11} for a review). 
The theoretical understanding, acquired from the analysis of the Boltzmann equation 
\cite{CastroNeto09,Peres2010}, numerical simulations 
\cite{Lewenkopf08,Mucciolo10,Lewenkopf13}, and field theoretical techniques 
\cite{Ostrovsky2006} is that extrinsic disorder such as ad-atoms absorbed in 
the graphene surface, substrate charge inhomogeneities, and intrinsic disorder 
such as vacancies play a key role. 

Motivated by recent experiments \cite{Lundeberg10,Wada2010,Wakabayashi2012,Couto2014}, 
we study the effect of extrinsic and intrinsic random gauge potential disorder due to strain.
We show that, although unlikely to be dominant in graphene deposited over standard 
substrates, these kinds of disorder give unique and sizable contributions to the conductivity.
For high-quality substrates, there are even experimental evidences \cite{Couto2014} that random 
strain can be a good candidate for the leading electron relaxation mechanism in on-substrate 
graphene.

Standard electronic transport experiments use samples where graphene flakes 
are deposited over an insulating substrate. In such setting, it has been experimentally 
established that the graphene surface is characterized by disordered static ripples 
\cite{Ishigami07,Geringer09,Deshpande09,Cullen10}. For SiO$_2$ substrates, the latter 
have typical lengths of $\lambda\approx 5 - 30$ nm and characteristic heights of  
$h_{\rm rms}\approx 0.2 - 0.5$ nm.   

Lattice deformations due to ripples change the distance between the atoms in the graphene
sheet. At the quantum level, site lattice displacements change the orbital bonding between 
the corresponding atoms, modifying the electronic structure of the material. In graphene, 
whose low-energy electronic properties are nicely described by a nearest-neighbor tight-binding 
model\cite{CastroNeto09}, the occurrence of ripples change the tight-binding hopping 
terms\cite{CastroNeto09,Kim2008,deJuan12}. For distortions with length scales much
larger than the lattice parameter, characteristic to most samples \cite{Ishigami07,Geringer09,Deshpande09,Cullen10}, 
it has been shown that the tight-binding model can be mapped into an effective Dirac Hamiltonian 
with a pseudo-magnetic vector potential \cite{Guinea08,Vozmediano12,Vozmediano10,CastroNeto09} 
that depends on the lattice distortions. Hence, random ripples give rise to an {\it intrinsic} random 
gauge potential. The experimental evidences of pseudo-magnetic fields are scarce and indirect, but 
quite remarkable. Strain fields have been invoked to associate the local density of states observed 
in graphene nanobubbles \cite{Levy2010} to Landau levels with energies corresponding to very high 
magnetic fields. To the best of our knowledge, no transport experiment has yet observed manifestations of 
this physical picture.

Random magnetic fields can also be achieved by realizing that disordered ripples in graphene 
and the roughness at semiconductor interface heterostructures share several common features. 
Starting at the late 80's, a number of ingenious methods where used to characterize the interface 
roughness in the heterostructures \cite{Mensz87,Anderson93,Zumbuhl04}. One idea is particularly 
suited for graphene studies: By applying a strong magnetic field $B_\parallel$, aligned to the plane of a
heterostructure interface confining the two-dimensional electron gas (2DEG), the (smooth) interface 
roughness disorder gives origin to a local random magnetic field perpendicular to the 2DEG surface. 
The analysis of the electronic transport properties as a function of the applied magnetic field gives 
quantitative information about the interface roughness.

This setting was nicely explored by a recent experiment\cite{Lundeberg10}, that combined information of the 
average conductivity and its weak localization correction \cite{McCann06,Kechedzhi2007} in graphene to extract 
the sample characteristic $\lambda$ and $h_{\rm rms}$. This procedure have been also used in the experimental 
study of other graphene systems \cite{Wada2010,Wakabayashi2012}.
Theory \cite{Mathur01} shows that an applied $B_\parallel$ on  a rough surface gives rise to an effective dephasing 
$\ell_\phi$ and to the suppression of the weak localization peak. In addition to this quantum correction, the random 
magnetic field due to $B_\parallel$ also contributes to the electron momentum relaxation, which at high doping is 
accounted for by the Boltzmann theory \cite{Lundeberg10}. 
This nice analysis does not consider the effect of intrinsic pseudo-magnetic fields due to strain, discussed 
above.  


Our focus is different. We study the combined effect of intrinsic and extrinsic random magnetic fields 
in the Drude conductivity. 
We revisit the analysis of the Boltzmann equation in graphene \cite{CastroNeto09,Novikov07,Mucciolo10,DasSarma11}
and calculate the contributions of random magnetic fields to the Drude conductivity. We show that the 
conductivity corrections due to an applied in-plane magnetic field $B_\parallel$ on a rippled graphene 
flake depend on the direction of $B_\parallel$ and are very anisotropic. We find that this result can be
reconciled with the experimental findings of Ref.~\onlinecite{Lundeberg10} by theoretically treating the 
effect of strain and $B_\parallel$ at the same footing. We also show that the combined effect of both sources 
of random magnetic fields provides a new experimental path to quantitatively probe the effects of strain
fields in the low-energy electronic dynamics of graphene.

The paper is structured as follows: In Section \ref{sec:model} we present the disorder model we employ 
to describe ripples and the corresponding effective Hamiltonian. We briefly explain the origin of an effective 
random field due to an applied in-plane magnetic field $B_\parallel$, as well as from the intrinsic strain field due to ripples. 
In Sec.~\ref{sec:theory} we calculate the contributions to the Drude conductivity due to $B_\parallel$ and 
strain using the Boltzmann transport equation. We show how to account for the disorder potential anisotropy 
and discuss its consequences comparing with experiments.
Finally, in Sec.~\ref{sec:conclusions} we present our conclusions and an outlook.

\section{model hamiltonian} 
\label{sec:model}

In this section we present a model to study the effect of extrinsic and intrinsic 
sources of a random magnetic fields in the dynamics of electrons in corrugated 
graphene monolayer samples. 

Close to the charge neutrality point, the electronic dispersion relation of pristine graphene 
monolayers is linear and has two degenerate components, with corresponding $K$ 
and $K'$ valley indices\cite{CastroNeto09}.
In the presence of a magnetic field, the effective electronic Hamiltonian for the $K$-valley 
reads 
\be 
\label{eq:effectiveDiracH}
H^{K} = v_F {\bm \sigma} \cdot [{\bm p} + e {\bm A}({\bm r})] = H^{K}_0 + V({\bm r}),
\ee
where the vector potential $\bm A(\bm r)$ has been included in $H^{K}$ by minimal coupling. 
Here $v_F\approx 10^{6}$m/s, $\bm \sigma$ are the Pauli matrices acting on the sublattice 
space, and $\bm p$ is the electron momentum operator. The Hamiltonian for the $K'$ valley 
has a similar structure \cite{CastroNeto09}.

In this description, a generic long-ranged disorder potential $V({\bm r})$  is represented in both 
$K$ and $K'$ valleys by
\be
\label{eq:disorderpotential}
V({\bm r}) =\sum_i \sigma_i V^{(i)}({\bm r}),
\ee
where $i=0$ stands for scalar disorder (with $\sigma_0=I_2$) and $i=1,2$ for vector potential 
disorder, while $i=3$ represents a mass term. The focus of our study are (intrinsic and extrinsic) 
disordered gauge fields, associated with $V^{(1)}$ and $V^{(2)}$. In this paper we do not consider 
scalar disorder.

Let us introduce 
\begin{align}
\label{eq:Veff}
\left\langle \bm{k}'s' |V| \bm{k}s \right\rangle & = 
\frac{1}{2} \left[ 1 + ss'e^{i(\theta-\theta')} \right] V^{(0)}_{{\bm k} - {\bm k}'} + 
\\ & \hskip-1.0cm
 \frac{s e^{i\theta}}{2} \!\left (V^{(1)}_{{\bm k} - {\bm k}'} -i V^{(2)}_{{\bm k} - {\bm k}'}\right) 
+\frac{s'e^{-i\theta'}}{2} \!\left(V^{(1)}_{{\bm k} - {\bm k}'} +i V^{(2)}_{{\bm k} - {\bm k}'}\right) \!
,
\nonumber
\end{align}
where the spinor
\begin{align}
\label{base}
\left|\bm{k}s\right\rangle &= \frac{1}{\sqrt{2\mathcal{A}}} 
\left(\begin{array}{c}
1 \\ s e^{i\theta}
\end{array}\right)
e^{i\bm{k}\cdot \bm r},
\end{align}
is an eigenstate of $H^K_0$,  $\theta=\tan^{-1}(k_y/k_x)$, $s$ indicates particle ($s=+1$) 
or hole ($s=-1$) doping, and 
\be
V^{(i)}_{{\bm k} - {\bm k}'} = \frac{1}{\cal A} \int \! d{\bm r} \, e^{i({\bm k} - {\bm k}')\cdot {\bm r}} 
\, V^{(i)} ({\bm r})
\ee
is the momentum representation of $V^{(i)}$. Since we deal with elastic processes, we assume in
the remaining of the paper that $s=s'$.

At low temperatures, scalar disorder (short and long ranged) is the main source of momentum 
relaxation in graphene systems \cite{Mucciolo10,DasSarma11}. In this paper we use a 
phenomenological transport time $\tau_s$ to account for effects of scalar disorder in the 
conductivity. We assume that $\tau_s$ is much shorter than the characteristic transport 
times due to random gauge fields. In Sec.~\ref{sec:conclusions}, where we compare our 
results to experiments, we show that $\tau_s$ indeed dominates the conductivity in 
graphene, but some transport properties are only explained by including effects due to 
random gauge fields.

The ripple disorder model employed in this study is defined as follows: We describe 
the graphene sheet surface by $z=h(\bm r)$, where $h$ is the surface displacement 
with respect to the reference plane $z=0$ at the position $\bm r =(x,y)$. The average 
of $h$ is set to zero. In line with the experiments on graphene deposited over a 
substrate\cite{Lundeberg10,Ishigami07,Geringer09,Deshpande09,Cullen10}, we 
further assume that the typical heights $h_{\rm rms}$ are much smaller that the ripple 
lengths $\lambda$.

We model the ripple fluctuations in $h(\bm r)$  by the correlation function 
\begin{equation}
\langle h({\bm r}) h({\bm r} ')\rangle = h_{\rm rms}^2\, F \!\left( \frac{ |{\bm r} - {\bm r}' |}{\lambda}\right),
\label{eq:rmsF}
\end{equation}
where $\langle \cdots \rangle$ denotes an average over disorder. Although theory predicts
a power law height-height correlation function for free-standing membranes \cite{Katsnelson2008}, 
experiments support single-parameter correlations for the (static) ripples of graphene deposited 
over a substrate. For latter  convenience, let us define 
\begin{equation}
h({\bm q})=\frac{1}{\mathcal{A}}\int d\bm r e^{i\bm q \cdot \bm r}h({\bm r}),
\end{equation}
where $\mathcal{A}$ is the sample size. In reciprocal space 
\begin{equation}
\langle h({\bm q}) h({\bm q'})  \rangle = h_{\rm rms}^2\ \overline{F}(\bm q)\delta_{\bm q, -\bm q'},
\label{hhcorrelatorfourierspace}
\end{equation}
where $\overline{F}(\bm q)$ is the Fourier transform of the correlation function $F(|\bm r- \bm r'|)$.

We address two mechanisms that generate random magnetic fields. First, we study the case of an 
external strong magnetic field $\bm B_{\parallel}$ applied parallel to the graphene sheet. We show 
that, due to the ripples, $\bm B_{\parallel}$ gives rise to a random effective magnetic field 
$B_{\rm ext}({\bm r})$  perpendicular to the graphene surface. Next, we discuss the intrinsic 
pseudo-magnetic field $\bm B_{\rm int}$ originated by the strain field corresponding to the graphene 
sheet profile height $h(\bm r)$.

\subsection{Random magnetic field due to ripples and an in-plane external B-field} 
\label{sec:geometry}

Let us first consider the setup of a magnetic field applied parallel to the sample $z=0$, 
that has been experimentally investigated in a variety of systems 
\cite{Mensz87,Anderson93,Zumbuhl04,Lundeberg10}. For notational convenience, in 
what follows we fix the direction of $B_{\parallel}$ along the $x$-axis, namely, 
$\bm B_{\parallel}=B_{\parallel}\hat{\bm x}$.

\begin{figure}[htbp]
\centering \includegraphics[width=0.9\columnwidth]{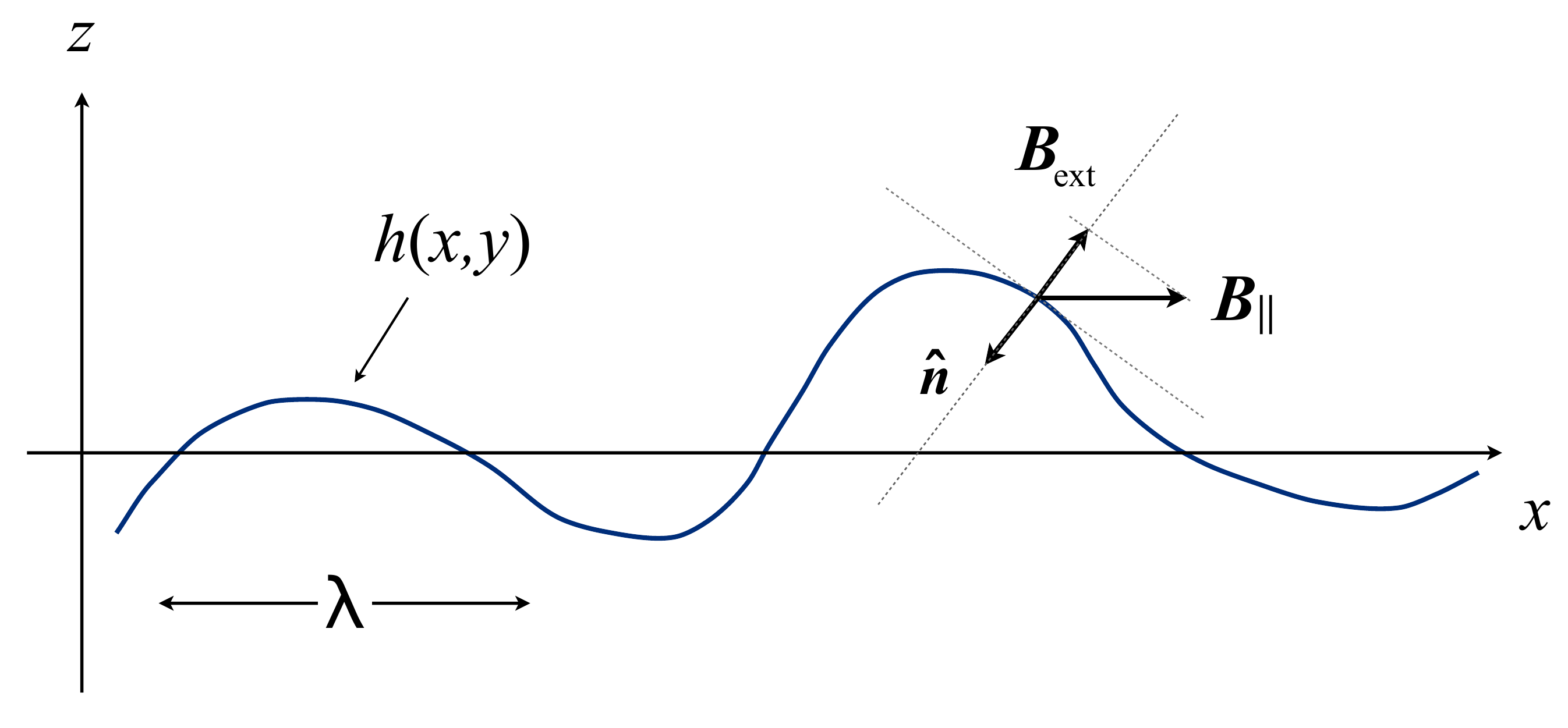}
\caption{Sketch of $h({\bm r})$ along the $x$ direction. The ripple amplitudes 
$\delta h$ are enhanced and made comparable with $\lambda$ to help the 
illustration. For convenience, we take $\bm B_{\parallel}=B_{\parallel}\hat{\bm x}$.}
\label{fig:h_xy}
\end{figure}

As illustrated in Fig.~\ref{fig:h_xy}, the parallel magnetic field ${\bm B}_\parallel$ has a 
component perpendicular to the surface $z=h({\bm r})$ that is given by 
\begin{align}
	B_{\rm ext} ({\bm r}) = 
	-{\bm B}_\parallel \cdot \hat{\bm n}({\bm r}).
\end{align}

At the point ${\bm r}_0=(x_0, y_0)$, the surface $z= h({\bm r})$ has a unit normal vector 
\begin{align}
	\hat{\bm n}({\bm r}_0) 
	= \frac{1}{ \sqrt{ 1 + 
	\left(\frac{\partial h}{\partial x}\right)^2 + 
	\left(\frac{\partial h}{\partial y}\right)^2
	} } 
	\left.\left(\begin{array}{c}
	\partial h/\partial x \\ \partial h/\partial y \\ -1 \end{array}\right)\right|_{{\bm r}={\bm r}_0}.
\end{align}
We assume that the typical displacement magnitude is characterized by $\delta h$. 
For $\delta h \ll \lambda$, we write
 \be
  \label{normalvector2}
\hat{\bm n}({\bm r}_0) \approx (
 \partial h/\partial x , \partial h/\partial y , -1 )\Big|_{{\bm r}={\bm r}_0}.
 \ee
Hence, the effective local perpendicular magnetic field reads
 \be
 \label{externalmagneticfield}
 B_{\rm ext} ({\bm r}) =  -{\bm B}_\parallel \cdot {\bm \nabla} h({\bm r}),
 \ee 
and is expressed, in a convenient gauge for $\bm B_{\parallel}=B_{\parallel}\hat{\bm x}$, 
by the vector potential
\begin{equation}
\label{eq:BparA}
A_x({\bm r}) = 0 \quad \mbox{and} \quad  A_y({\bm r}) = -B_{\|}h(\bm r)\,.
\end{equation}

\begin{figure}[htbp]
\begin{tabular}{cc}
\includegraphics[width=0.85\columnwidth]{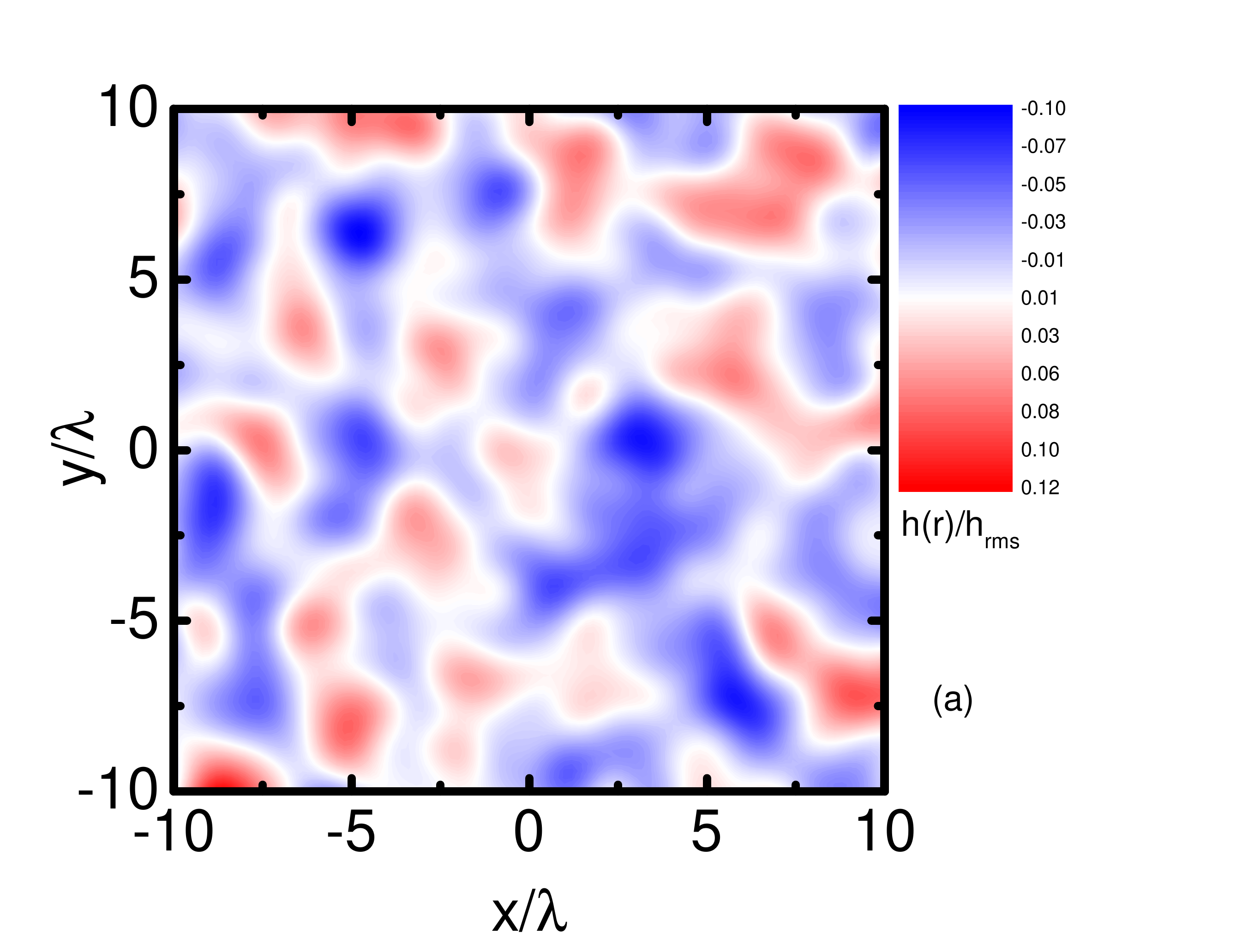}\\
\includegraphics[width=0.85\columnwidth]{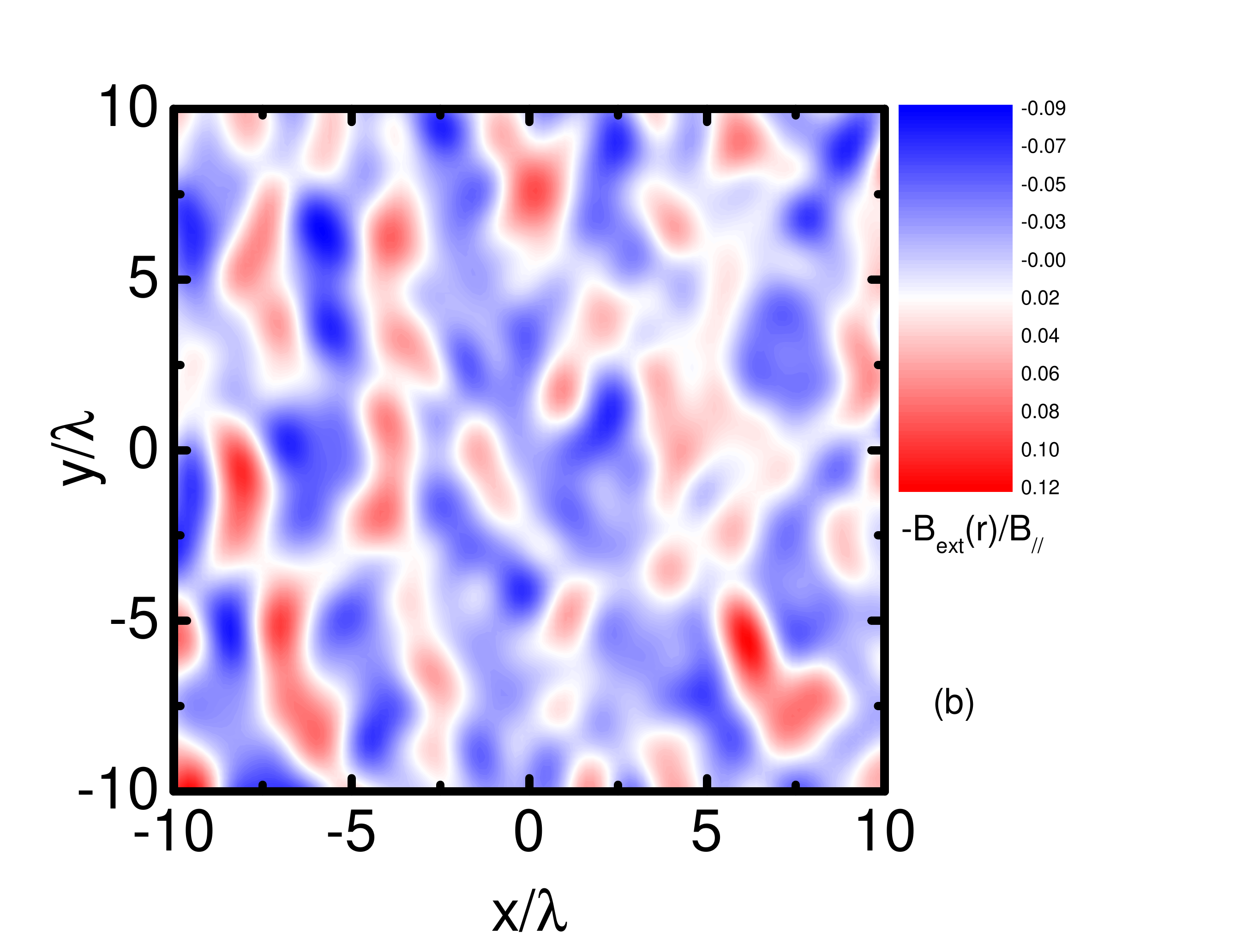} \\
\includegraphics[width=0.85\columnwidth]{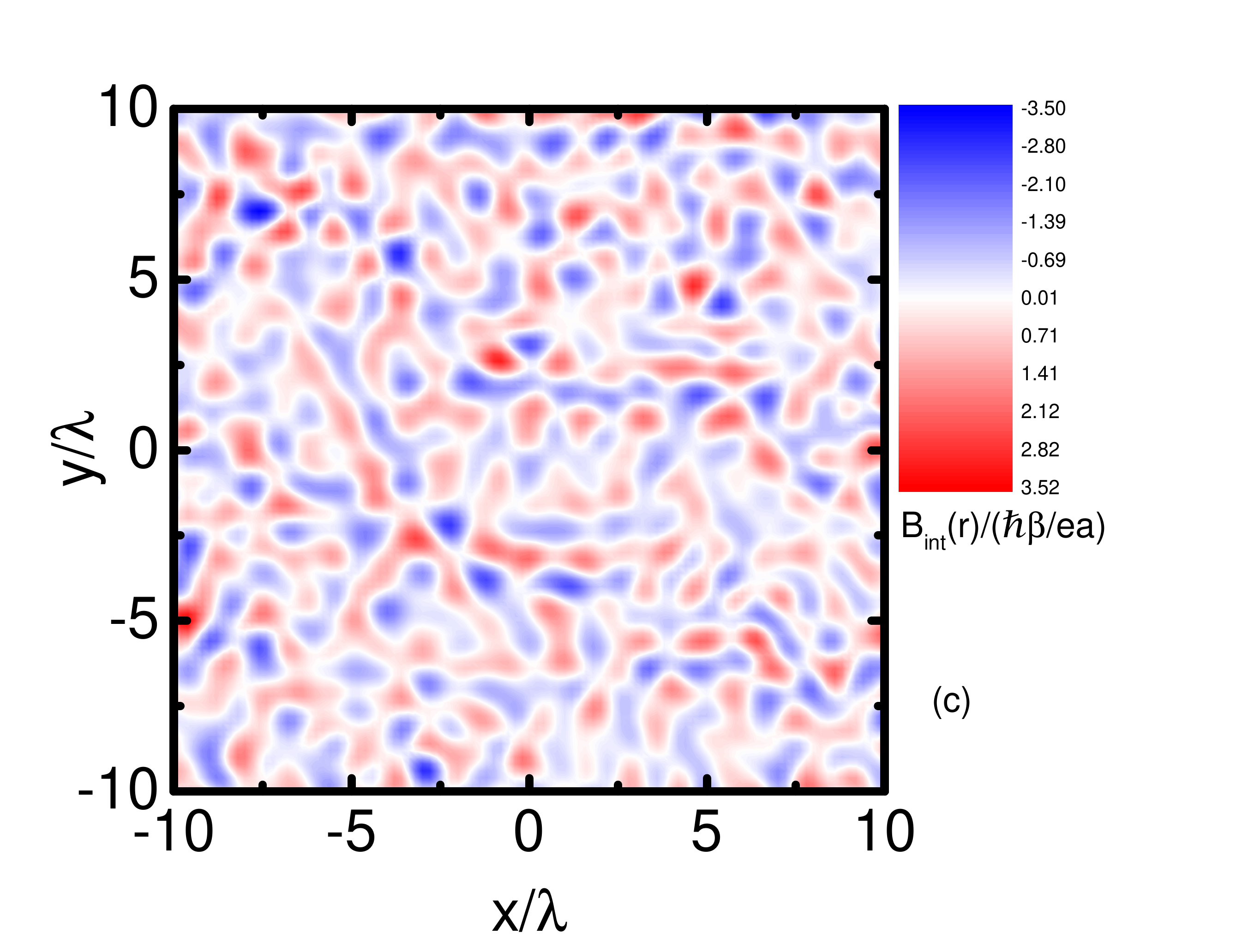} 
\end{tabular}
\caption{(color online) Typical disorder realization of (a) $h({\bm r})$, characterized by a 
Gaussian correlation function $F$, the corresponding (b) $B_{\rm ext}({\bm r})$ 
for an external ${\bm B}_\parallel$ applied along the $x$ direction, defined in 
Eq.~\eqref{externalmagneticfield}, and (c) $B_{\rm int}({\bm r})$ due to lattice 
deformations, given by Eq.~\eqref{strainvectorpotential}.}
\label{fig:random_fields}
\end{figure}

Figure \ref{fig:random_fields}(a) illustrates a typical disorder realization of $h(\bm r)$ with 
fluctuations characterized by the Gaussian correlation function $F(x)=\exp{(-x^2/2\lambda^2)}$. 
The corresponding magnetic field $B_{\rm ext} ({\bm r})$, normal to the graphene sheet, is 
shown in Fig.~\ref{fig:random_fields}(b). 
While $h(\bm r)$ displays an isotropic disorder, $B_{\rm ext} ({\bm r})$ is clearly anisotropic. 
The anisotropy direction of $B_{\rm ext} ({\bm r})$ depends on the orientation of $B_{\parallel}$. 

The anisotropy is quantified by inspecting the autocorrelation function 
\begin{equation}
\label{autocorrelationfuncition}
\langle B_{\rm ext}(\bm r) B_{\rm ext}(\bm r')\rangle=B^2_{\parallel}
\left\langle \frac{\partial h(\bm r)}{\partial x} \frac{\partial h(\bm r')}{\partial x'}\right\rangle,
\end{equation}
that can be expressed in terms of $F$ by direct differentiation. Alternatively, 
going to reciprocal space, one writes
\begin{eqnarray}
\left\langle \frac{\partial h(\bm r)}{\partial x} \frac{\partial h(\bm r')}{\partial x'}\right\rangle&=&
h^2_{\rm rms}\sum_{\bm q}q^2_x \, \overline {F}(\bm q)e^{-i\bm q \cdot(\bm r -\bm r')}\nonumber\\
&=&-h^2_{\rm rms}\frac{d^2}{dx^2} F(\bm \rho),
\end{eqnarray}
with $\bm \rho=\bm r - \bm r'$.

\begin{figure}[htbp]
\vskip0.2cm
\begin{tabular}{cc}
\includegraphics[width=0.85\columnwidth]{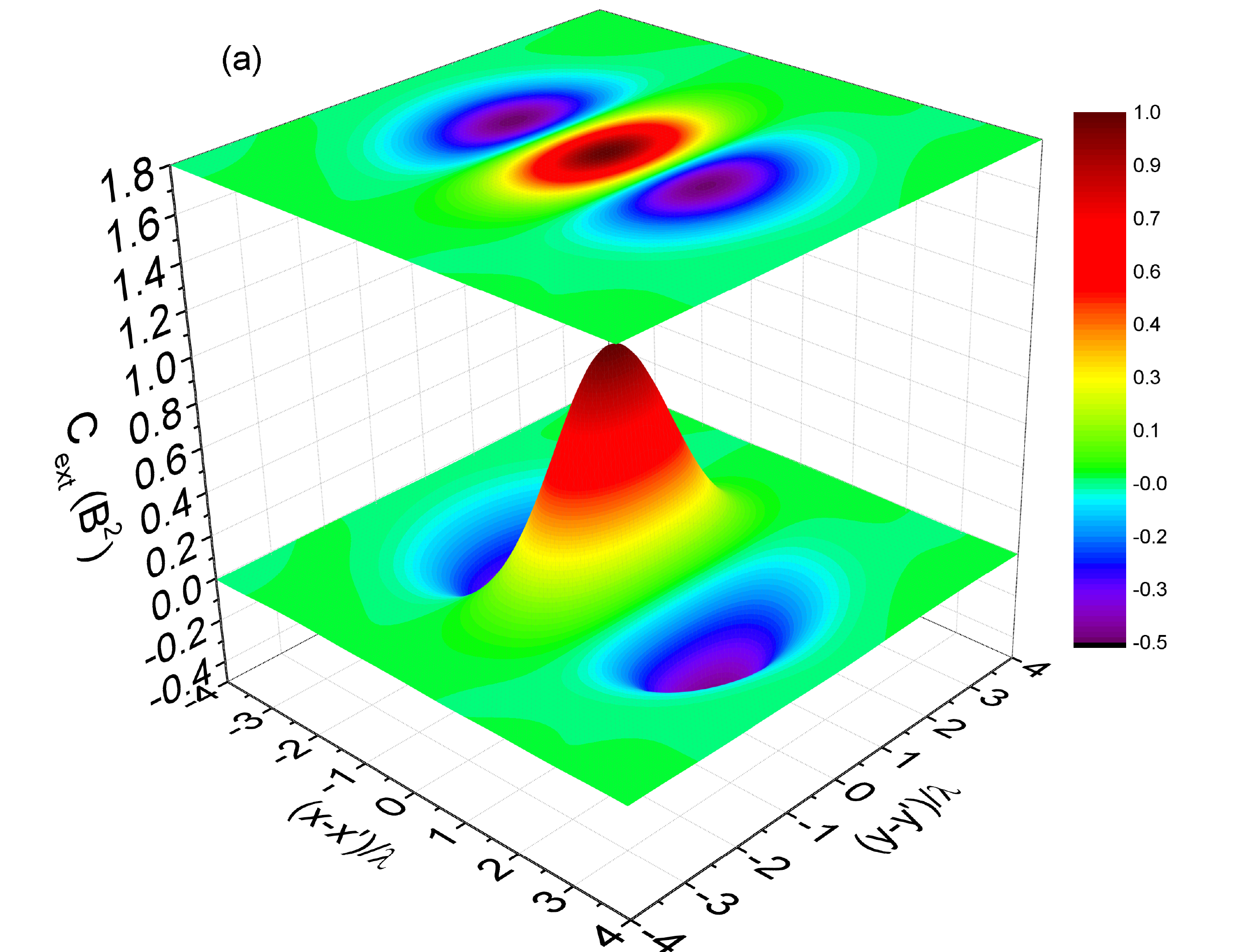}\\
\includegraphics[width=0.85\columnwidth]{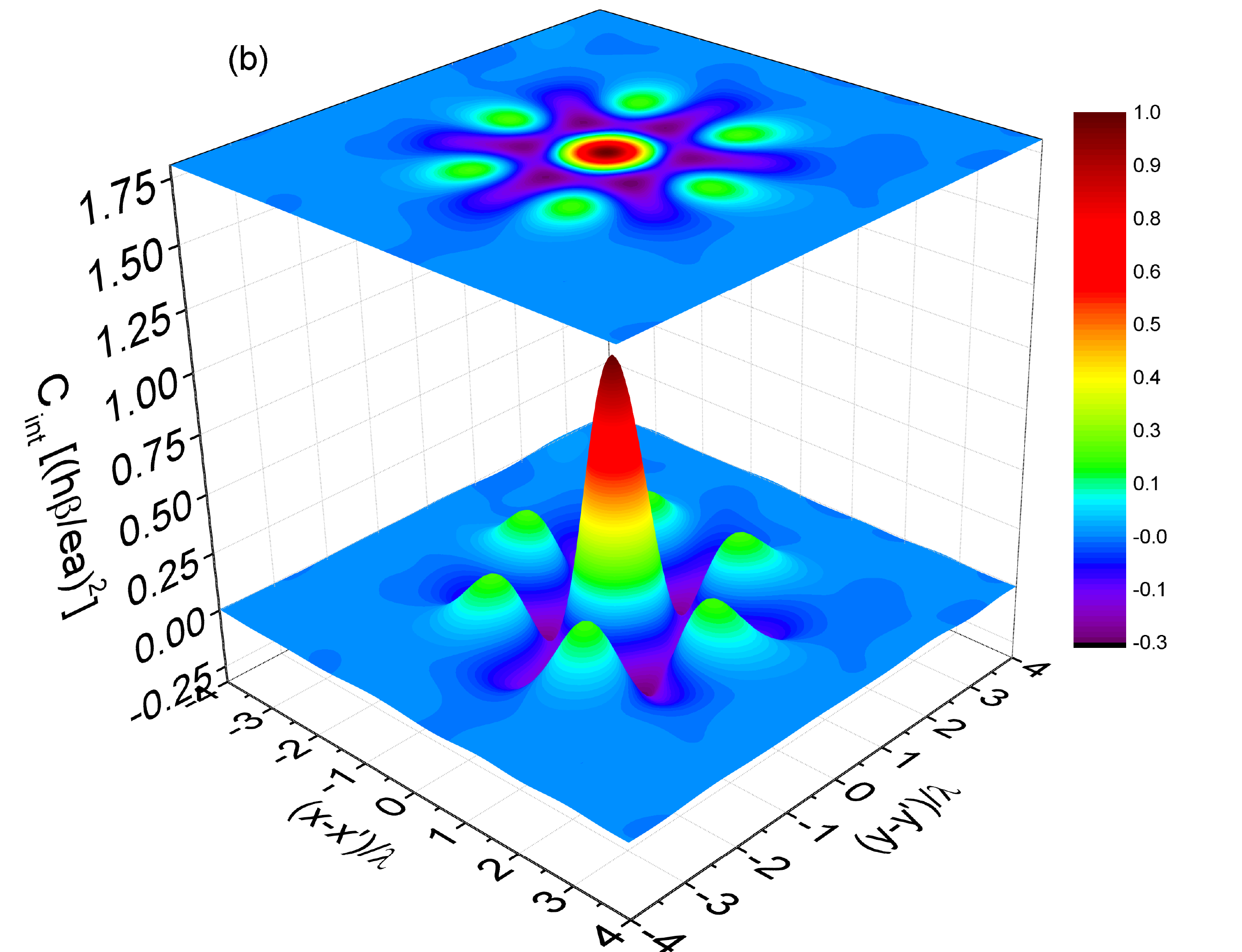}
\end{tabular}
\caption{(color online) Extrinsic and intrinsic magnetic field correlations functions:
(a) $C_{\rm ext}({\bm r} - {\bm r}') = \langle B_{\rm ext}({\bm r}) B_{\rm ext}({\bm r}')\rangle$ 
for an external ${\bm B}_{\parallel}$ applied along the $x$-direction and 
(b) $C_{\rm int}({\bm r} - {\bm r}') =\langle B_{\rm int}({\bm r})B_{\rm int}({\bm r}')\rangle$ due
 to strain in units of $h_{\rm rms}^4/\lambda^6(\hbar \beta/e a)^2$, both 
 corresponding to a ripple disordered surface $h (\textbf{r})$ characterized 
 by a Gaussian correlation function.}
\label{fig:correlators}
\end{figure}

Let us calculate $\langle B_{\rm ext}(\bm r) B_{\rm ext}(\bm r')\rangle$ for the case 
where $h({\bm r})$ is characterized by a Gaussian correlation function, the same as
in Fig.~\ref{fig:random_fields}(a). The corresponding $B_{\rm ext}(\bm r)$ autocorrelation 
function reads
\begin{equation}
\label{correlatorexternalfield}
\langle B_{\rm ext}(\bm r) B_{\rm ext}(\bm r')\rangle
=B^2_\parallel 
\frac{h_{\rm rms}^2}{\lambda^2}
\left[1-\left(\frac{\rho}{\lambda}\right)^2 \! \cos^2\alpha\right]
e^{-\frac{\rho^2}{2 \lambda^2}}\\
\end{equation}
where $\alpha$ is the angle between $\bm B_{\parallel}$ (or the $x$-axis) and $\bm \rho$. 

Figure \ref{fig:correlators}(a) shows the $B_{\rm ext}(\bm r)$ autocorrelation function 
obtained by averaging over $10^5$ ripple disorder realizations of  $h({\bm r})$, as defined 
by Eq.~\eqref{eq:rmsF} with a Gaussian correlation function $F$ (for more details see, for 
instance, Ref.~\onlinecite{Lewenkopf08}).
As expected, it coincides with Eq.~\eqref{correlatorexternalfield} and expresses the 
anisotropy captured by a visual inspection of Fig.~\ref{fig:random_fields}(b). 


\subsection{Pseudomagnetic field due to strain} 
\label{sec:pseudo}

The out of plane deformations of a rippled membrane described by $h(\bm r)$ can be associated 
with the strain tensor $u_{ij}(\bm r)$ given by \cite{Ando2002,Manes2007,Guinea08}
\begin{align}
u_{xx} & \approx \frac{1}{2}\!\left(\frac{\partial h}{\partial x}\right)^2, \quad
u_{yy}    \approx \frac{1}{2}\!\left(\frac{\partial h}{\partial y}\right)^2, \quad \mbox{and}
\nonumber\\
u_{xy} & \approx \frac{1}{2}\!\left(\frac{\partial h}{\partial x}\frac{\partial h}{\partial y}\right).
\end{align}
For simplicity we have neglected the effect of in-plane deformations.

The effect of strain in the low-energy electronic structure of graphene can be accounted 
for by introducing a scalar and a vector gauge potential in the Dirac equation \cite{Katsnelson2008,Vozmediano10,Vozmediano12}.
The scalar term reads
\begin{equation}
\label{strainscalarpotential}
V^{(0)}(\bm r)=g\left[u_{xx}(\bm r)+u_{yy}(\bm r)\right],
\end{equation}
while, for the $K$-valley and for an armchair crystallographic orientation along the $x$-axis,  
${\bm A} = (A_x, A_y)$ is given by
\begin{align}
\label{strainvectorpotential}
A_x(\bm r)&=\frac{\hbar\beta\kappa}{ea}[u_{xx}(\bm r)-u_{yy}(\bm r)], \nonumber\\
A_y(\bm r)&=-2\frac{\hbar\beta\kappa}{ea} u_{xy}(\bm r) \,,
\end{align}
where $a=1.42$ \AA \ is the bond length between nearest neighboring carbon atoms, 
and $e$ is the electron charge. Here $g\approx 4$ eV, $\kappa\approx 1/3$, and $\beta=-\partial \log t/\partial
 \log a \approx 2$ are dimensionless material dependent parameters\cite{Guinea08,Ando2002} that 
characterize the coupling between the Dirac electrons and the lattice deformations, where $t \approx 3$ eV 
is the hopping integral between nearest-neighbor $\pi$-orbitals.
 
In summary, for any given $h(\bm r)$ one can readily calculate the pseudo-magnetic field 
$\bm B_{\rm int}=\nabla\times \bm A$. Since $A_z=0$ and neither $A_x$ nor $A_y$ depend 
on $z$, $\bm B_{\rm int}=B_{\rm int}\hat{\bm z}$. 
Figure \ref{fig:random_fields}(c) shows the $B_{\rm int}$ corresponding to the random 
rippled surface $h(\bm r)$ of Fig. \ref{fig:random_fields}(a).
Notice that the typical correlation length of $B_{\rm int}({\bm r})$ is much shorter than that of $h(\bm r)$.

Let us calculate $\langle  B_{\rm int}(\bm r)  B_{\rm int}(\bm r')\rangle$ for a random Gaussian 
correlated $h(\bm r)$, corresponding to Eq.~\eqref{eq:rmsF} with $F(x)=e^{-x^2/2\lambda^2}$. 
To this end, we calculate the Fourier transform of the intrinsic pseudo magnetic field, namely
\be
\label{Fouriercampostrain}
 B_{\rm int}(\bm q)=
i\frac{\hbar \beta\kappa}{e a}
[q_{y}u_{xx}(\bm q) +2q_{x}u_{xy}(\bm q)-q_{y}u_{yy}(\bm q)],
\ee
with
\begin{equation}
\label{Fouriertensorstrain}
u_{ij}(\bm q)=-\frac{1}{2} \sum_{\bm q'}q'_{i}(q_{j}-q'_{j}) h(\bm q')h(\bm q-\bm q'), 
\end{equation}
where $i$ and $j$ label the Cartesian coordinates. 

We use Eqs.~\eqref{Fouriercampostrain} and \eqref{Fouriertensorstrain} to write the correlation function of 
$B_{\rm int}$ in momentum space. The evaluation of $\langle B_{\rm int}(\bm q) B_{\rm int}(-\bm q)\rangle$ 
amounts to compute the corresponding $\langle u_{ij}^{}(\bm q)u_{i'j'}^{}(-\bm q)\rangle$, that result in four-$h$
correlation functions. This can be done exactly for Gaussian fluctuations and provides a good qualitative estimate 
for other cases\cite{Katsnelson2008}.

\begin{widetext}
We obtain 
\begin{eqnarray}
\label{momentumspacecorrelatorstrain}
\langle B_{\rm int}(\bm q) B_{\rm int}(-\bm q)\rangle
&=&
\frac{h_{\rm rms}^4 \pi}{32  \lambda^2 \mathcal{A}}\left(\frac{\hbar \beta \kappa}{e a}\right)^2
q^2\Bigg[
16+\lambda^4q^4
\sin^23\theta
\Bigg]e^{-\lambda^2q^2/4},
\end{eqnarray}
where $\theta$ is the angle between $\bm q$ and the $x$-direction.

By Fourier transforming back to coordinate space, we arrive at
\be
\label{realspacecorrelatorstrain}
C_{\rm int}({\bm r} - {\bm r}') \equiv
\langle  B_{\rm int}(\bm r)  B_{\rm int}(\bm r')\rangle
=
\frac{h_{\rm rms}^4}{\lambda^6}
\left(\frac{\hbar \beta \kappa}{e a}\right)^2
\Bigg[
8
-20\frac{\rho^2}{\lambda^{2}}
+9\frac{\rho^4}{\lambda^{4}}
-2\frac{\rho^6}{\lambda^{6}}\sin^2 3\alpha
\Bigg]e^{-\rho^2/\lambda^2},
\ee
\end{widetext}
where $\alpha$ is the angle between ${\bm \rho} = {\bm r} - {\bm r}'$ and the $x$-axis. 
 \footnote{
Since we only considered the out of plane contribution in the strain tensor, 
Eq.~\eqref{momentumspacecorrelatorstrain} gives a slightly different correlation 
function than the one found in Ref.~\onlinecite{Guinea08}, where a more general 
expression for the strain tensor was used.}

The correlation function $C_{\rm int}({\bm r} - {\bm r}')$ has 6 symmetry axes, 
reflecting the underlying graphene honeycomb lattice symmetry\cite{Guinea08}. In other words, 
information about the graphene crystal structure survives disorder averaging. 
Figure~\ref{fig:correlators}(b) shows 
$C_{\rm int}({\bm r} - {\bm r}')$ obtained from $10^5$ numerical realizations of Gaussian 
correlated disorder for $h({\bm r})$. The numerical simulations serve as a helpful test to check 
our analytical results. As in the previous subsection, we verify an excellent agreement within the
statistical precision.
 
\section{Drude-Boltzmann conductivity} 
\label{sec:theory}

In this Section we use the effective Dirac Hamiltonian of Eq.~\eqref{eq:effectiveDiracH}
to calculate the transport time and the Drude-Boltzmann conductivity of  graphene 
monolayers in the presence of random gauge fields. 

High mobility graphene samples have typical electronic mean free paths of $\ell \gtrsim 50$ nm. 
Recalling \cite{CastroNeto09} that the carrier density is related to the Fermi wave number by 
$k_F = \sqrt{\pi |n|}$, one readily obtains that $k_F \ell \gg 1$ already for a doping where
$|n| \approx 10^{11}$ cm$^{-2}$. This indicates that even for modest carrier densities 
a semiclassical transport description is justified. For $|n| \gtrsim 10^{11}$ cm$^{-2}$ 
the typical graphene conductivity in good samples is much larger than $e^2/h$, the order of
magnitude of quantum contributions to the electronic transport, such as weak localization 
\cite{McCann06,Kechedzhi2007,Mucciolo10} and universal conductance fluctuations 
\cite{Hinz2014}. In such situations, the Boltzmann approach is very successful in assessing 
the conductivity, as shown by direct comparison with numerical simulations using an atomistic 
basis \cite{Lewenkopf08,Klos2010}. As one approaches the charge neutrality point, and 
$k_F \ell \lesssim 1$, the semiclassical method is no longer suited and one has to resort to 
more sophisticated approaches \cite{Ostrovsky2006,Mucciolo10}.

We now discuss how to add gauge field disorder in the Boltzmann approach.
For long ranged disorder, some authors \cite{Kittel1987,Hedegard95,Brandao2014} 
argue that it can be advantageous to include the disorder potential in the classical 
Liouvillian evolution, that is, to treat $V({\bm r})$ in the {\it left-hand side} of the 
Boltzmann equation. 
This approach is justified in the ``classical" regime, where $k_F\lambda\gg 1$, 
that is, where the random fields with a characteristic length $\lambda$ vary 
slowly in the scale of $k_F^{-1}$.
In graphene samples with ripple sizes $\lambda$ of the order of few to ten nanometers 
\cite{Ishigami07,Geringer09,Deshpande09,Cullen10}, 
the latter inequality holds for a carrier density $|n|\gg 10^{12}$ cm$^{-2}$, 
that is much larger than the doping studied in most experiments \cite{CastroNeto09}.

In this study, we calculate transport times for both short and long ranged disorder by 
evaluating the corresponding Boltzmann collision integral 
(at the {\it right-hand side} of the equation). As mentioned in the introduction, for graphene 
on standard substrates (the case of interest here), the random gauge field contribution to 
the conductivity is not the dominant one. Hence, the electron mean free path due to ripples 
is larger than $\ell$ and the arguments justifying the semiclassical approximation hold.

The Boltzmann equation for graphene under a uniform electric field $\bm E$ reads
\cite{Ziman72, CastroNeto09}
\begin{equation}
\label{Boltzmann}
-e \bm E \cdot \frac{\partial \varepsilon_{\bm k, s}}{\partial \bm p}
\frac{\partial f_0}{\partial \varepsilon} =
\sum_{{\bm k'},s'}(g_{\bm k,s}-g_{\bm k',s'})\mathcal{W}_{{\bm k'},s \leftarrow {\bm k},s}
\end{equation}
where $\varepsilon_{\bm k, s}=sv_F\hbar|\bm k|$, $f_0$ is the Fermi distribution function, 
$g_{\bm k}=f_{\bm k}-f_{0}$ is deviation from equilibrium due to the electric field, and 
$W_{{\bm k'},s' \leftarrow {\bm k},s}$ is the transition rate from state $({\bm k},s)$ to $({\bm k}', s)$,
 that we calculate using Fermi golden rule, namely
\begin{equation}
\label{eq:FGR}
\mathcal{W}_{{\bm k}',s' \leftarrow {\bm k},s} =
 \frac{2\pi}{\hbar} 
\left \langle  |\langle\bm{k}'s'|V|\bm{k}s\rangle|^2 \right\rangle
 \delta(\varepsilon_{\bm k,s} - \varepsilon_{{\bm k}',s'}),
\end{equation}
where $V$ is a generic long-ranged disorder potential parametrized by 
Eq.~\eqref{eq:disorderpotential}. The $\delta$-function reflects the fact we are dealing 
with elastic processes and, hence, $s=s'$. In our model, the transition rates do not 
depend on $s$. Accordingly we drop this index whenever its omission does not introduce 
an ambiguity.

The scattering processes we address are anisotropic. The calculation of the transport 
properties in this case is slightly different \cite{Herring55,Sondheimer62} than that of
the standard isotropic case \cite{Ziman72}. In this study, we adapt the nice method developed 
by Tokura \cite{Tokura1998} -- that is briefly described in what follows -- to calculate the
transport times of massless Dirac electrons in graphene.

In both situation of interest, the scattering potential correlation functions have at least one 
symmetry axis. For convenience, we choose the $x$-axis along a symmetry axis and 
define \cite{Tokura1998}
\begin{equation}
\label{perturbation}
g_{\bm k}=\left(-\frac{\partial f_0}{\partial \varepsilon_{\bm k}}\right)e v_{\bm k} \,
                 {\bm \tau}(\theta)\cdot \bm E,
\end{equation}
where $\bm \tau(\theta)$ is the relaxation time vector to be solved. We recall that 
$\theta$ is the angle between ${\bm k}$ and the $x$-axis. Note that ${\bm \tau}$ 
depends explicitly on $\theta$ and implicitly on $|\bm k|$.

The current density (spin and valley degeneracies included) is given by
\begin{align}
\bm j&=\frac{4}{\mathcal{A}}\sum_{\bm k}e \bm v_{\bm k} g_{\bm k}
\nonumber\\
&=\frac{e^2}{\pi^2}\int^{\infty}_0 \! dk k \left(-\frac{\partial f_0}{\partial \varepsilon_{\bm k}}\right)\!
\int^{2\pi}_0 \! d \theta \, v_{\bm k} \, \bm v_{\bm k} [\bm \tau(\theta)\cdot\bm E],
\end{align}
from which one obtains the conductivity tensor
\begin{equation}
\label{eq:conductivity_tensor}
\sigma=\frac{e^2|\varepsilon_F|}{\hbar^2\pi^2}
\int^{2\pi}_0d \theta
\left(
\begin{array}{cc}
 \tau_x(\theta)\cos\theta & \tau_y(\theta)\cos\theta   \\
 \tau_x(\theta)\sin\theta  & \tau_y(\theta)\sin\theta  
\end{array}
\right),
\end{equation}
where $\varepsilon_F$ is the Fermi energy, measured with respect to the charge neutrality 
point energy. For the sake of simplicity, in Eq.~\eqref{eq:conductivity_tensor} we 
have taken the zero-temperature limit, namely, 
$-\partial f_0/\partial \varepsilon=\delta(\varepsilon-\varepsilon_F)$.

By substituting the ansatz \eqref{perturbation} in the Boltzmann equation, Eq.~\eqref{Boltzmann}, 
one obtains an integral equation for $\bm \tau(\theta)$, namely
\begin{eqnarray}
\label{taox}
\cos\theta&=&\int^{2\pi}_{0}d\theta'[\tau_x(\theta)-\tau_x(\theta')]\cal{W}(\theta, \theta')
\\
\label{taoy}
\sin\theta&=&\int^{2\pi}_{0}d\theta'[\tau_y(\theta)-\tau_y(\theta')]\cal{W}(\theta, \theta'),
\end{eqnarray}
where $\theta'$ is the angle between $\bm k'$ and the $x$-axis, and
\begin{align}
\label{eq:angularW}
{\cal{W}}(\theta, \theta')&=
\frac{{\cal A}}{(2\pi)^2}\int^{\infty}_0 \!dk' k'  \mathcal{W}_{{\bm k}',s \leftarrow {\bm k},s}
\nonumber\\
&=
\frac{{\cal A} |\varepsilon_F|}{2\pi v^2_F \hbar^3}
\left\langle|\langle {\bm k}',s |V |{\bm k}, s \rangle|^2 \right\rangle,
\end{align}
where, due to the zero-temperature limit,  $k=k'=k_F$.

The matrix element $\langle {\bm k}',s |V |{\bm k}, s \rangle$ depends on 
$\bm q=\bm k-\bm k'$ and $\varphi=\pi/2+(\theta+\theta')/2$, the angle between 
$\bm q$ and the $x$-axis. Hence, ${\cal{W}}(\theta, \theta')$ is
better cast as ${\cal{W}}(q_\zeta, \varphi)$.
We use the standard notation $\zeta=|\theta- \theta'|$ and $q_{\zeta}=2k_F\sin(\zeta/2)$.

By expressing $\tau_x$ and $\tau_y$ in terms of a Fourier series, one transforms 
Eqs.~\eqref{taox} and \eqref{taoy} into an (infinite) set of algebraic equations. Using 
the $x$-axis symmetry and that ${\cal{W}(\theta,\theta')}={\cal{W}(\theta',\theta)}$
\footnote{Microreversibility is usually invoked to guarantee ${\cal{W}(\theta,\theta')}=
{\cal{W}(\theta',\theta)}$ and Eq.~\eqref{Boltzmann}. In the presence of an external 
magnetic field, that breaks time-reversal symmetry, ${\cal{W}(\theta,\theta')}=
{\cal{W}(\theta',\theta)}$ is still true within the approximation used in Eq.~\eqref{eq:FGR}}, 
one shows that\cite{Tokura1998}
\begin{eqnarray}
\label{taoxfourier}
\tau_x(\theta) & = & \sum_{n=1}^{\infty}\tau^{(n)}_{x}\cos[(2n-1)\theta] \\
\label{taoyfourier}
\tau_y(\theta) & = & \sum_{n=1}^{\infty}\tau^{(n)}_{y}\sin[(2n-1)\theta] .
\end{eqnarray}
By inserting the above relations in Eq.~\eqref{eq:conductivity_tensor}, one concludes 
that the conductivity tensor is diagonal, with
\begin{equation}
\sigma_{xx}=\frac{e^2|\varepsilon_F|}{\hbar^2\pi}\tau^{(1)}_{x} \quad {\rm and} \quad
\sigma_{yy}=\frac{e^2|\varepsilon_F|}{\hbar^2\pi}\tau^{(1)}_{y},
\end{equation}
that supports the interpretation of $\bm \tau^{(1)}$ as a transport time vector.

The symmetry ${\cal{W}(\theta,\theta')}={\cal{W}(\theta',\theta)}$ implies that 
${\cal{W}}(q_{\zeta},\varphi)={\cal{W}}(q_{\zeta},\varphi+\pi)={\cal{W}}(q_{\zeta},\varphi-\pi)$.
In turn
\begin{equation}
{\cal{W}}(q_{\zeta},\varphi)=\sum_{n=0}^{\infty}{\cal{W}}_{n}(q_{\zeta})\cos(2n\varphi),
\end{equation}
with an obvious inversion relation.

By using the Fourier expansions for $\bm \tau(\theta)$ and $W(q_{\zeta},\varphi)$, 
Eqs.~\eqref{taox} and \eqref{taoy} can be cast in matrix form \cite{Tokura1998} 
\be
\label{eq:matrixK}
\delta_{l,1} = \sum_{n=1}^\infty  M_{l, n}^-\tau_{x}^{(n)} \quad \mbox{and} \quad
\delta_{l,1} = \sum_{n=1}^\infty  M_{l, n}^+\tau_{y}^{(n)},
\ee
where the matrix elements of $M^{\pm}$ are\cite{Tokura1998}
\begin{equation}
M^{\pm}_{l,n}=\frac{(-1)^{l-n}}{2}\big[(1+\delta_{l,n})J_{|l-n|,n+l-1}\pm J_{n+l-1,|l-n|}\big],
\end{equation}
with
\begin{equation}
J_{n,m}=\int^{2\pi}_{0}d\zeta\,\mathcal{W}_n(q_{\zeta})[\cos (n\zeta)-\cos (m\zeta)].
\end{equation}
Finally, by inverting $M^{\pm}$ in Eq.~\eqref{eq:matrixK}, one writes the vector 
transport time components as
\begin{equation}
\tau^{(1)}_x=[(M^{-})^{-1}]_{11} \quad {\rm and} \quad \tau^{(1)}_y=[(M^{+})^{-1}]_{11}.
\end{equation}

Note that for isotropic scattering, all ${\cal{W}}_n$ with $n>0$ are zero and 
$M^{\pm}$ is diagonal, with elements $K_{l,l}=J_{0,2l-1}$. Hence, the vector 
transport time components coincide, $\tau^{(1)}_x=\tau^{(1)}_y=\tau^{(1)}$, 
and read
\begin{equation}
\frac{1}{\tau^{(1)}}=J_{0,1}=\int^{2\pi}_{0}d\zeta{\cal{W}}_{0}(q_{\zeta})(1-\cos\zeta),
\end{equation}
which is the standard expression for the transport time in isotropic systems.

\subsection{Effect of an in-plane magnetic field}
\label{sec:Bparallel_Boltzmann}

Let us now calculate the effect of an external parallel magnetic field on 
the conductivity. From Eq.~\eqref{eq:BparA} we write the effective disorder potential for 
the $K$-valley as
\be
V_{\rm ext}(\bm{r}) = 
v_F e \sigma_y A_y({\bm r}) = - v_F e B_\parallel h({\bm r}) \sigma_y.
\ee
We recall that $h(\bm r)$ varies slowly in the scale of the lattice spacing and, 
hence, $V_{\rm ext}(\bm r)$ is long-ranged and does not mix valleys.

The momentum relaxation rate $W_{{\bm k}' \leftarrow {\bm k}}$ reads
\be
{\cal{W}}_{{\bm k}' \leftarrow {\bm k}} =\delta(k-k') 
\frac{2\pi e^2v_F}{\hbar^2} B_\parallel^2
\sin^2\!\left(\frac{\theta + \theta'}{2}\right) \! \frac{C_h(q)}{\mathcal{A}}.
\ee
where 
\be
C_h(q) =\int \! d{\bm r} \,e^{i {\bm q} \cdot {\bm r}}\, 
\big\langle h(0) h({\bm r})\big\rangle
\ee 
is the form factor of the height-height correlation function. Here $\langle \cdots \rangle$ 
indicates disorder average. 

From Eq.~\eqref{eq:angularW} we obtain
\begin{equation}
\mathcal{W}(q,\varphi)=\frac{ (eB_\parallel)^2|\varepsilon_F|}{4 \pi \hbar^3} C_h (q)(1+\cos 2\varphi),
\end{equation}
that has only 2 non-zero Fourier components, namely,
\begin{equation}
\mathcal{W}_n(q)=\frac{ (eB_\parallel)^2|\varepsilon_F|}{4 \pi \hbar^3}  C_h (q), 
  \quad  \text{for} \quad n=0,1
\end{equation}
while $\mathcal{W}_n(q)=0$ for $n\geq 2$.

In this case,  the $M^{\pm}$ matrix is tridiagonal and reads \cite{Tokura1998}
\begin{equation}
\label{matrixcampoexterno}
M^{\pm}=
\begin{pmatrix}
  (1\mp \frac{1}{2}) J_{0,1}  & -\frac{1}{2}J_{1,2}   &  0  &  \cdots\\
    -\frac{1}{2}J_{1,2}  &   J_{0,3}  & -\frac{1}{2}J_{1,4}  & \cdots \\
   0              &   -\frac{1}{2}J_{1,4}  &     J_{0,5}             &   \cdots \\
     \vdots                  &     \vdots  &\vdots   &\ddots 
\end{pmatrix}.
\end{equation}
The inverse transport time components are given by
\begin{equation}
\label{inverseanisotropictime}
\frac{1}{\tau^{(1)}_{x}}=\frac{3}{2}J_{0,1}-\Gamma_3 
\quad \mbox{and} \quad
\frac{1}{\tau^{(1)}_{y}}=\frac{1}{2}J_{0,1}-\Gamma_3 ,
\end{equation}
where  $\Gamma_3$ is isotropic and determined by the 
continued fraction relation
\begin{equation}
\label{gammafactor}
\Gamma_m=\frac{( J_{1,m-1})^2}{4(J_{0,m}-\Gamma_{m+2})}.
\end{equation}
In practice, we compute $\Gamma_3$ by assuming that $\Gamma_{\overline{m}} =0$ 
and subsequent iteration of \eqref{gammafactor}. The choice of $\overline{m}$ determines
the precision of the calculation: The larger $\overline{m}$, the more accurate is $\Gamma_3$.

Assuming that $\Gamma_3\ll J_{0,1}$ leads to an interesting result, that is
\begin{equation}
\label{inversetimext0}
\frac{1}{\tau_y^{(1)}}=\frac{1}{3\tau_x^{(1)}}=
\frac{(eB_{\parallel})^2|\varepsilon_F|}{8\pi\hbar^3}\!
\int_0^{2\pi} \!d \zeta (1-\cos \zeta) C_h(q).
\end{equation}
In this limit $\tau_y^{(1)}/\tau_x^{(1)}=3$. In other words, for $\Gamma_3\ll J_{0,1}$ the 
corrections to the conductivity due to ${\bm B}_\parallel$ lead to $\Delta\sigma_{yy}=
3\Delta\sigma_{xx}$, regardless of the dependence of correlation function $C_h(q)$ on $q$. 

For $\Gamma_3=0$ and $C_h(q)=2\pi\lambda^2h_{\rm rms}^2 e^{-\lambda^2q^2/2}$,  we 
write $\tau_{x,y}^{(1)}$ in closed analytical form, namely
\begin{align}
\label{eq:tau_Gamma3=0}
\dfrac{1}{\tau_y^{(1)}}=
\frac{1}{3\tau_x^{(1)}}=&
\frac{(eB_{\parallel})^2|\varepsilon_F|}
{4\hbar^3}(\lambda h_{\rm rms})^2 
\nonumber\\ &\times e^{-\lambda^2k_F^2}\!
\left[I_0(\lambda^2 k_F^2)-I_1(\lambda^2 k_F^2) \right],
\end{align}
where $I_0$ and $I_1$ are modified Bessel functions of the first kind. We use \cite{CastroNeto09} 
$k_F=\sqrt{\pi|n|}$  to express the conductivity in terms of the charge carrier density $n$.
We conclude that  the correction to the conductivity due to an in-plane magnetic field depends
quadratically on $h_{\rm rms} B_\parallel$ and has a non-trivial dependence on $\lambda^2 |n|$. 

In the high doping limit of $\lambda |n|^{1/2} \gg1$, Eq.~\eqref{eq:tau_Gamma3=0} gives
the resistivity contribution of an in-plane magnetic field in a rippled graphene sheet
\begin{equation}
\label{eq:Delta_rho_as}
\Delta \rho_{yy}=3\Delta \rho_{xx}\approx \frac{1}{ 2\sqrt{2} \hbar} 
      \frac{h^2_{\rm rms}}{\lambda} B^2_{\parallel} |n|^{-3/2}
\end{equation}
in agreement with Ref.~\onlinecite{Lundeberg10}. 

\begin{figure}[htbp]
\centering \includegraphics[width=0.85\columnwidth]{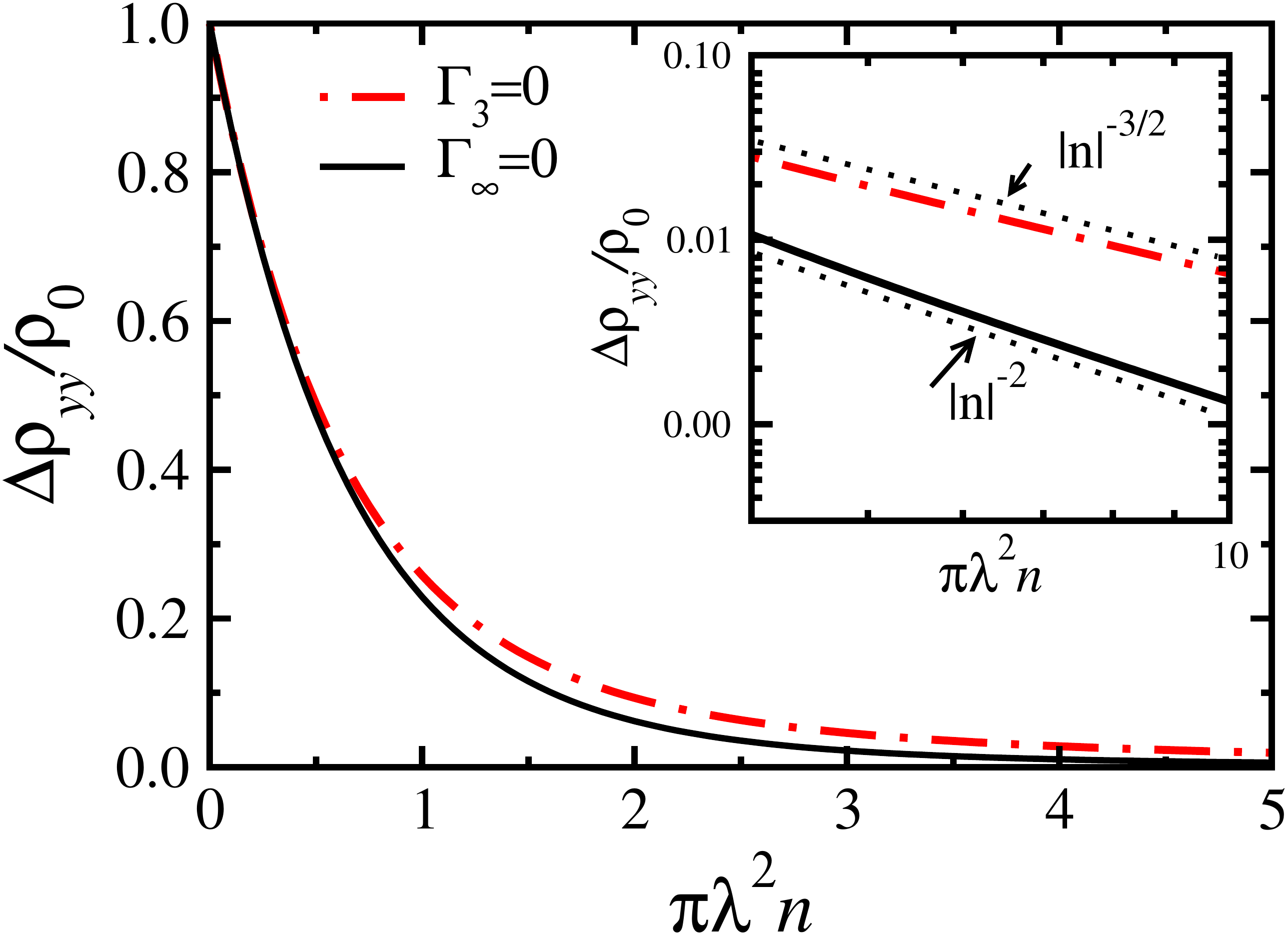}
\caption{(color online) Resistivity $\Delta \rho_{yy}$ in units of $\rho_0=(\pi\lambda 
h_{\rm rms}B_{\parallel})^2/2 \hbar$ due to ${\bm B}_\parallel$ as a function of $\lambda^2 n$. 
Inset: The same as in the main plot in log-log scale to illustrate the dependence of 
$\Delta \rho_{xx}$ on $|n|$.}
\label{conductivityparallel}
\end{figure}

Figure \ref{conductivityparallel} shows the resistivity $\Delta \rho_{yy}$ versus the carrier density $n$ 
(due to particle hole symmetry, we only show $n>0$) for $\overline m=3$ and $\overline m \rightarrow \infty$. 
The optimal $\overline m$ value to obtain convergence depends on $\lambda^2n$. 
The inset shows $\Delta \rho_{yy}$ for $\lambda^2n$ values outside the validity range of the asymptotic 
expansion. As discussed in the next section, the $\lambda^2n$ range displayed in the inset corresponds 
to the typical experimental situation. We find that the $|n|^{-3/2}$ scaling predicted by the asymptotic 
expansion \eqref{eq:Delta_rho_as} is only a rough approximation.

In general, $\Gamma_3$ is a non-vanishing correction to the transport time components, hence
$\tau_y^{(1)}/\tau_x^{(1)}\neq 3$. However, for Gaussian ripple height correlations, the ratio
$\tau_y^{(1)}/\tau_x^{(1)}$ is a function only of $\lambda k_F$. 

\begin{figure}[htbp]
\centering \includegraphics[width=0.85\columnwidth]{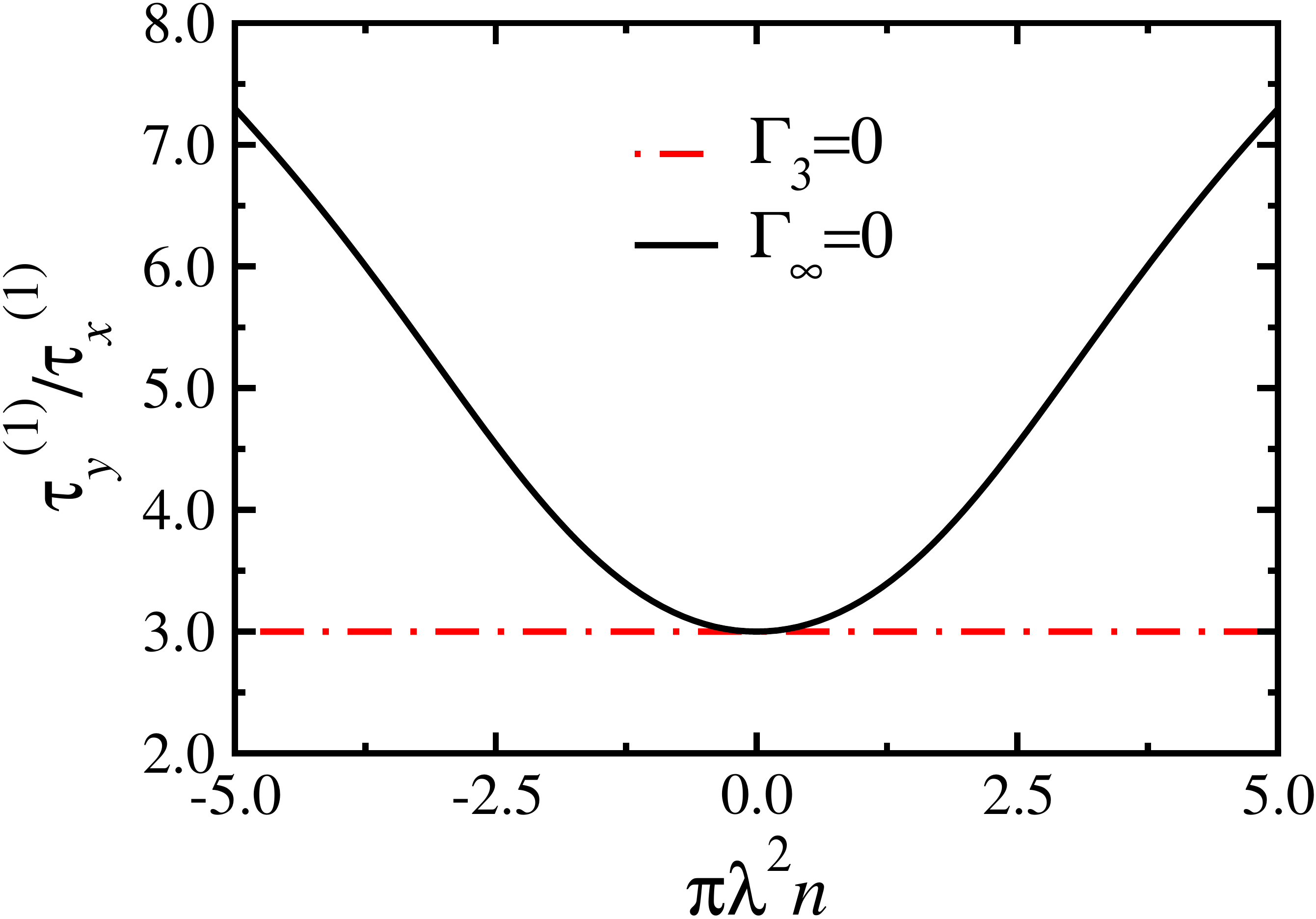}
\caption{(color online) Ratio $\tau_y^{(1)}/\tau_x^{(1)}$ as a function of $\pi\lambda^2n$. 
For $\pi\lambda^2n<5.0$, the limit $\overline{m} \rightarrow \infty$ is attained within $10^{-4}$ 
accuracy for $\overline{m}=13$.  
}
\label{fig:anisotropy}
\end{figure}

Figure \ref{fig:anisotropy} shows $\tau_y^{(1)}/\tau_x^{(1)}$ versus $\lambda^2 n$. It illustrates the
importance of $\Gamma_3$ in the calculation of the conductivity corrections. We find that by 
increasing the carrier concentration the anisotropic conductivity is considerably favored. 

The transport properties predicted by Eq.~(\ref{eq:tau_Gamma3=0}) are only slightly 
modified for the case of exponential ripple heigh correlations, $\langle h(0) h({\bm r})\rangle 
= h_{\rm rms}^2 e^{-r/\lambda}$: For $\lambda |n|^{1/2} \gg 1$, the resistivity tensor given 
by Eq.~\eqref{eq:Delta_rho_as} is multiplied  \cite{Genma2012} by a prefactor of order of 
unity times $\log(\lambda^2 |n|)$.

%
\subsection{Effect of strain fields}
 \label{sec:strain_Boltzmann}

Let us now consider the effect of the pseudo-magnetic fields due to strain, 
$B_{\rm int}$, in the conductivity of monolayer graphene sheets. 
In contrast to the mechanism discussed above, $B_{\rm int}({\bm r})$ is solely 
determined by $h({\bm r})$ and the material properties. Hence, it intrinsic to any 
graphene sample with disordered ripples.
 
We use Eq.~\eqref{strainvectorpotential} to calculate the vector transport time 
${\bm \tau}$ for the intrinsic effective vector potential due to strain.
For the $K$-valley $V_{\rm int}$ reads
\be
V_{\rm int}({\bm r}) = \hbar v_F \frac{\beta \kappa}{a} \Big\{
\left[u_{xx}({\bm r}) - u_{yy}({\bm r})\right]\sigma_x - 2 u_{xy}({\bm r}) \sigma_y \Big\}.
\ee

In contrast with the previous subsection, here it is difficult to make quantitative progress 
without assuming a specific form for the ripple height correlation function. The qualitative
behavior of the conductivity corrections due to strain that has been reported in the literature 
\cite{Katsnelson2008} is not sufficient for the analysis we propose.

As in Sec.~\ref{sec:pseudo}, we calculate $\langle u_{ij} ({\bm r}) u_{i'j'}({\bm r}')\rangle$  
by assuming Gaussian correlated ripple height fluctuations, 
$C_h(q)=2\pi\lambda^2h_{\rm rms}^2e^{-\lambda^2q^2/2}$. After some lengthy but  
straightforward algebra, we obtain
\begin{align}
\langle| \langle\bm{k}'s|V_{\rm int} & |\bm{k}s\rangle|^2\rangle= 
\frac{v_F^2\hbar ^2 \beta^2\kappa^2}{32a^2}
\frac{\pi h_{\rm rms}^4 }{ \lambda^2 \mathcal{A}}
\\ 
&\times\bigg\{ 
16+\lambda^4 q^4\cos^2\!\left[\frac{3}{2}(\theta+\theta')\right]
\bigg\}e^{-\lambda^2q^2/4}.
\nonumber
\end{align}
Using Eq.~\eqref{eq:angularW} we arrive at
\be
{\cal{W}}(q, \varphi)= {\cal W}_{\rm int}
\left[16+ \frac{\lambda^4 q^4}{2}(1+\cos6\varphi) \right]e^{-\lambda^2q^2/4},
\end{equation}
where 
\be
 {\cal W}_{\rm int} =  \frac{\beta^2 \kappa^2  |\varepsilon_F|}{64a^2\hbar} \frac{h_{\rm rms}^4}{\lambda^2}.
\ee
(The notation is the same as that of the previous section.) 

The only non-zero Fourier components of ${\cal{W}}(q, \varphi)$ are
\begin{eqnarray}
{\cal{W}}_{0}(q)&=& {\cal W}_{\rm int}\left(16+\frac{\lambda^4 q^4}{2}\right)
e^{-\lambda^2q^2/4} \quad \mbox{and} 
\nonumber\\
{\cal{W}}_{3}(q)&=&{\cal W}_{\rm int}
\frac{\lambda^4 q^4}{2}e^{-\lambda^2q^2/4}.
\end{eqnarray}

In this case, the $M^{\pm}$ matrix reads\cite{Tokura1998}
\begin{equation}
\label{matrixcampointrinsic}
M^{\pm}=
\begin{pmatrix}
   J_{0,1}   &                          0                          &  \mp \frac{1}{2}J_{3,2} & -\frac{1}{2}J_{3,4} & \cdots\\
    0           &    J_{0,3} \mp \frac{1}{2}J_{3,0}  &                  0                   &             0               &\cdots \\
   \mp \frac{1}{2}J_{3,2}  &  0                           &              J_{0,5}             &   0 &  \cdots \\
   -\frac{1}{2}J_{3,4}          &  0  &  0                           & J_{0,7} & \cdots \\
     \vdots                  &     \vdots  &\vdots   &\vdots  &\ddots 
\end{pmatrix},
\end{equation}
and the inverse transport time components are given by
\begin{equation}
\label{inversetimestrain}
\frac{1}{\tau^{(1)}_{x/y}}=
J_{0,1}-\frac{(J_{3,2})^2}{4J_{0,5}}-\frac{(J_{3,4})^2}{4J_{0,7}}+\cdots.
\end{equation}
Since $\tau^{(1)}_{x}=\tau^{(1)}_{y}$, the conductivity corrections due 
to the strain field are isotropic. This result seems to be at odds with the fact that the 
pseudo-magnetic field autocorrelation function $\langle B_{\rm int}({\bm r}) 
B_{\rm int}({\bm r}')\rangle$ clearly shows an hexagonal symmetry, as 
illustrated by Fig.~\ref{fig:correlators}b. As shown by Tokura \cite{Tokura1998}, using
general arguments, this is a false paradox: The conductivity tensor becomes 
anisotropic only for scattering processes characterized by a single symmetry axis, 
like in the $B_\parallel$ case, analyzed in the previous subsection.

Assuming that $J_{0,1}$ vastly dominates the sum in 
Eq.~\eqref{inversetimestrain}, we obtain 
\begin{widetext}
\be
\label{eq:tau_strain_Gauss}
\frac{1}{\tau^{(1)}_{x/y}}= 
{\cal W}_{\rm int}
\pi e^{-\lambda^2k^2_F/2}
\Big[
(32+8\lambda^2k^2_F+16\lambda^4k^4_F)I_{0}(\lambda^2k^2_F/2)
-(64+24\lambda^2k^2_F+16\lambda^4k^4_F)I_{1}(\lambda^2k^2_F/2)
\Big],
\ee
\end{widetext}
where $I_0$ and $I_1$ are modified Bessel functions of the first kind. 

In the limit of $\lambda |n|^{-1}\gg1$, we obtain the correction to the resistivity
\begin{equation}
\Delta\rho_{xx}=\Delta\rho_{yy}\approx
 \frac{h}{e^2} \frac{23 \beta^2 \kappa^2}{32 \pi}  \frac{h_{\rm rms}^4}{\lambda^2 a^2}
\lambda^{-3} |n|^{-3/2}.
\end{equation}
The above asymptotic leading order expansion for $\Delta \rho$ helps us to develop 
some insight on the relevant parameters. However, since the situation of $k_F \lambda\gg 1$ 
is hardly met in the current experimental situations of interest, it is necessary to numerically 
calculate the inverse transport times, as we do in the next subsection. 
As expected the strain corrections to the conductivity depend on material parameters, 
and are a non-trivial function of $\lambda$, $h_{\rm rms}$, and $|n|$. Since these corrections
are small compared to other disorder effects, they are difficult to be noticed in standard 
transport experiments. This situation changes if we consider the combined effect of intrinsic 
and extrinsic random magnetic fields, as we discuss in the next. 

\subsection{Combined effect of extrinsic and intrinsic random magnetic fields}
 \label{sec:combination}

We conclude this section by analyzing the combined effect of both previously discussed
sources of random magnetic field disorder.  
As before, we assume that the system transport 
properties are dominated by other scattering processes, with a corresponding (isotropic) 
transport time $\tau_{\rm s}$. 
 
It is customary to use Matthiessen's rule when dealing with systems characterized by
different competing relaxation time mechanisms.   
In our case, Matthiessen's rule translates into adding the inverse transport times 
given by  Eqs.~\eqref{inverseanisotropictime} and \eqref{inversetimestrain}, namely
\begin{equation}
\label{inversetimeMR}
\frac{1}{\tau^{(1)}_{x/y}}=\frac{2\pm1}{2}J^{\rm ext}_{0,1}-\Gamma^{\rm ext}_3
+
J^{\rm int}_{0,1}-\frac{(J^{\rm int}_{3,2})^2}{4J^{\rm int}_{0,5}}-
\frac{(J^{\rm int}_{3,4})^2}{4J^{\rm int}_{0,7}}+\cdots .
\end{equation}
This naive approach was shown to be inaccurate when dealing with anisotropic 
potentials\cite{Tokura1998}. 

We analyze the combined effect of intrinsic and extrinsic random magnetic fields 
by considering an effective $M$-matrix given by
\begin{equation}
M^{\pm}_{\rm tot} =M^{\pm}_{\rm ext}+M^{\pm}_{\rm int}, 
\end{equation}
where $M^{\pm}_{\rm ext}$ and $M^{\pm}_{\rm int}$ are given by Eqs.~\eqref{matrixcampoexterno} 
and \eqref{matrixcampointrinsic} respectively. 

The ${\bm \tau}$ components are
\begin{equation}
\tau^{(1)}_{x}=[(M^{-}_{\rm tot})^{-1}]_{11} \quad {\rm and} \quad \tau^{(1)}_{y}=[(M^{+}_{\rm tot})^{-1}]_{11}, 
\end{equation}
that, for $\overline{m}=5$, explicitly read
\begin{widetext}
\begin{eqnarray}
\label{inversetimetotal}
\frac{1}{\tau^{(1)}_{x/y}}&=&
(1\pm \frac{1}{2}) J^{\rm ext}_{0,1}+J^{\rm int}_{0,1} 
+\frac{1}{(J^{\rm ext}_{0,3}+J^{\rm int}_{0,3}\mp J^{\rm int}_{3,0}/2)(J^{\rm ext}_{0,5}
+ J^{\rm int}_{0,5})-(J^{\rm ext}_{1,4})^2/4}
\left[
-\frac{1}{4}(J^{\rm ext}_{0,5}+ J^{\rm int}_{0,5})(J^{\rm ext}_{1,2})^2\right.
\nonumber\\
&& \left.\mp  \frac{1}{4}J^{\rm int}_{3,2}J^{\rm ext}_{1,2}J^{\rm ext}_{1,4} -
\frac{1}{4}(J^{\rm int}_{3,2})^2
(J^{\rm ext}_{0,3}+J^{\rm int}_{0,3}\mp \frac{1}{2}J^{\rm int}_{3,0})
\right]+\cdots.
\end{eqnarray} 
\end{widetext}
This result is clearly different from Eq.~(\ref{inversetimeMR}), since it mixes intrinsic and 
extrinsic effects. Let us now discuss the dependence of ${\bm \tau}$ on $B_\parallel, \lambda, 
h_{\rm rms}$, and $n$. For that purpose we numerically invert the matrix $M^\pm_{\rm tot}$, 
at order $\overline{m} \approx 30 - 50$  to guarantees an accuracy of $10^{-5}$ 
for the analyzed parameter range. 

\begin{figure}[htbp]
\centering{\includegraphics[width=0.85\columnwidth]{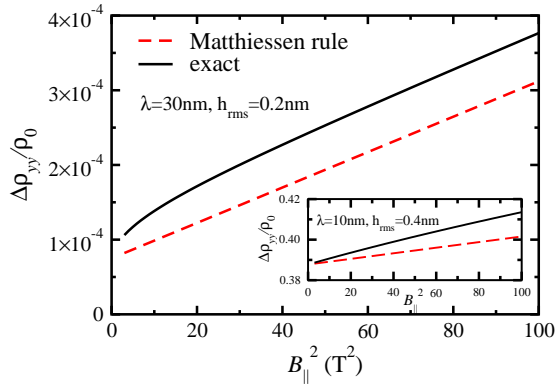}}
\caption{(color online) Resistivity correction $\Delta\rho_{yy}$ in units of $\rho_0=(\pi\lambda 
h_{\rm rms}B_{\parallel})^2/2 \hbar$ as a function of the in-plane magnetic field $B_\parallel$ for 
$h_{\rm rms}=0.2$ nm, $\lambda=30$ nm, and $n=10^{12}\,\text{cm}^{-2}$. Inset: The same 
as in the main figure for $h_{\rm rms}=0.4$ nm and $\lambda=10$ nm.}
\label{resistivity-vs-parameters}
\end{figure}

The resistivity corrections obtained from the Matthiessen's rule, Eq.~(\ref{inversetimeMR}) depend 
quadratically on $B_\parallel$, in line with the experiment \cite{Lundeberg10}. The full $M$-matrix analysis 
does not guarantee this simple dependence. In Fig.~\ref{resistivity-vs-parameters} we plot the resistivity correction 
$\Delta\rho_{yy}$ calculated using the full $M$-matrix and compare it with the one obtained from the 
Matthiessen rule, given by Eq.~(\ref{inversetimeMR}), for realistic values of $h_{\rm rms}$, $\lambda$, 
and $n$.
The full $M$-matrix calculation (indicated as ``exact") shows an overall higher resistivity than that 
obtained from the Matthiessen rule. It depends linearly on $B_\parallel^2$ for large $B_\parallel$ 
and deviates from this dependence only when 
$B_\parallel$ becomes small.

In contrast, the dependence of $\Delta \rho$ with $h_{\rm rms}$, $\lambda$, and $|n|$ is not trivial. 
For $h_{\rm rms}$ and $\lambda$  values taken close to the ones reported by topography experiments  
\cite{Ishigami07,Geringer09,Deshpande09,Cullen10}, 
a numerical study using the full $M$-matrix approach gives $\Delta\rho_{yy} \propto \lambda^{-\alpha}$ 
with $\alpha \approx 3 \cdots 4$, $\Delta\rho_{yy} \propto h_{\rm rms}^{\beta}$ with $\beta \approx 3$, and 
$\Delta\rho_{yy} \propto |n|^{-\gamma}$ with $\gamma \approx 2$.
In summary, $\Delta \rho$ is very sensitive on small variations of $h_{\rm rms}$ and $\lambda$.

\begin{figure}[htbp]
\centering \includegraphics[width=0.85\columnwidth]{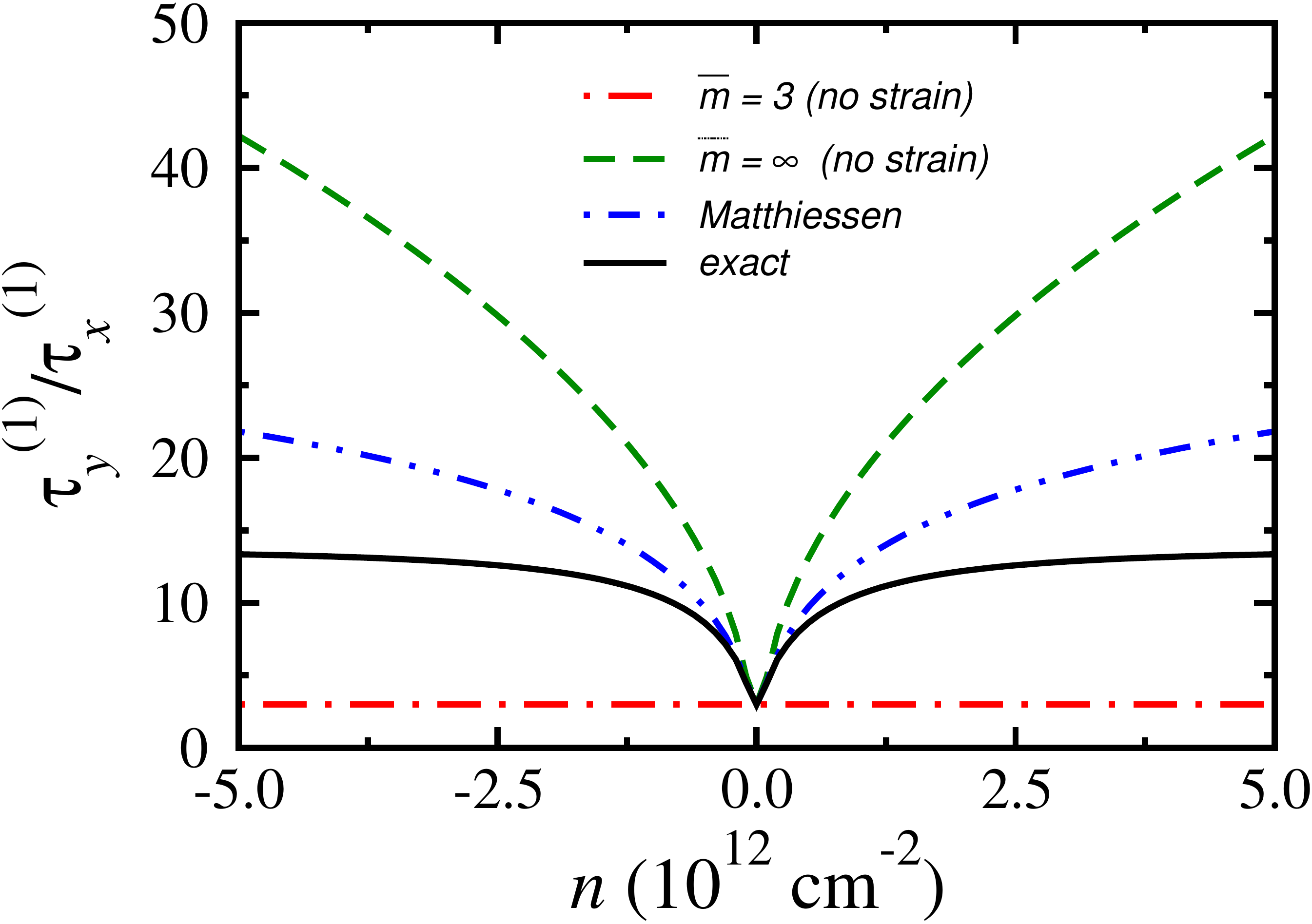}
\caption{(color online) Anisotropy $\tau_y^{(1)}/\tau_x^{(1)}$ as a function of the carrier 
concentration $n$ using different approximation schemes, for $h_{\rm rms}=0.2$ nm, 
$\lambda=30$ nm, and $B_{\parallel}=8$ T.  The red and green lines stand for the 
contribution of $B_\parallel$ without accounting for strain fields. The blue line represents 
the contributions of both external and strain fields using the conventional Matthiessen's 
rule. The black line stands for the combined effect of intrinsic and extrinsic fields obtained 
for the full $M$-matrix analysis. }
\label{fig:extvsintrinsic}
\end{figure}

In Fig.~\ref{fig:extvsintrinsic} we compare Eqs.~\eqref{inverseanisotropictime}, \eqref{inversetimeMR}, and
\eqref{inversetimetotal} to gain insight on how the strain mechanism affects the ratio $\tau_y^{(1)}/\tau_x^{(1)}$.
We find that the strain fields contribute to a strong suppression of the anisotropy in the transport time
due to a strong $B_\parallel$. However, for realistic parameter values the anisotropy is still very large and 
of the order of $\tau_y^{(1)}/\tau_x^{(1)} \approx 10$ for $|n| \approx 10^{12}$ cm$^{-2}$.

In order to further compare our results with the experiment\cite{Lundeberg10}, let us introduce the 
magnetoresistance $\Delta \rho={\bm E} \cdot {\bm j}/j^2$,
where $j_x=j \cos\xi $, $j_y=j \sin\xi$, and $\xi$ is the angle between $\bm B_{\parallel}$ and $\bm j$. 
Using the relation $E_i=\rho_{ij}J_j$ one writes \cite{Rushford2004}
\begin{equation}
\label{magnetoresistivity}
\Delta \rho(\xi) = \Delta\rho_{xx}\cos^2\xi+\Delta\rho_{yy}\sin^2\xi .
\end{equation}
Ref.~\onlinecite{Lundeberg10} reports $\Delta \rho(70^{\rm o})/\Delta \rho(20^{\rm o})\approx 0.13 - 0.26$. 
Using $\lambda$ and $h_{\rm rms}$ values obtained from AFM measurements, we obtain $\tau_y^{(1)}/\tau_x^{(1)} \approx 10$ 
for $n = 10^{12}$ cm$^{-2}$. This ratio leads to $\Delta \rho(70^{\rm o})/\Delta \rho(20^{\rm o}) \approx 0.2$ in good agreement
with the experiment \cite{Lundeberg10}.

\section{Conclusions and Outlook}
\label{sec:conclusions}

In this paper we studied the effect of random magnetic fields on the transport properties of a rippled 
graphene flake. We used the Boltzmann equation, adapted to the case of anisotropic disorder \cite{Tokura1998},
to address the case of an external magnetic field applied in-plane, the effect of 
intrinsic strain fields caused by the graphene corrugation, as well as the combination of both.

We find that an external in-plane magnetic field $B_\parallel$ gives rise to very anisotropic 
conductivity corrections. By neglecting the effect of strain fields and using a parametrization of the 
ripple disorder that is consistent with experiments, we find conductivity corrections that scale with 
$B_\parallel^2$ and $|n|^{-2}$, consistent with Ref.~\onlinecite{Lundeberg10}. In contrast, we obtain 
$\tau_y^{(1)}/\tau_x^{(1)}$ ratios as large as $20 \cdots 30$. 

In the absence of an external magnetic field, random gauge fields due to ripple disorder give a 
{\it isotropic} contribution to the electron momentum relaxation in graphene, in line with the order of 
magnitude estimate presented in Ref.~\onlinecite{Katsnelson2008}.
We find, however, that ripple disorder cannot be neglected 
in the analysis of the conductivity in the presence of a large $B_\parallel$. We also conclude that, due to the 
anisotropic nature of the problem, Matthiessen's rule is not accurate to address both intrinsic and extrinsic random 
fields at the same footing. For that purpose we have to invert the total $M$-matrix. 

This approach allows us to successfully describe the corrections to the Drude conductivity reported in
 the experiment \cite{Lundeberg10} using typical $\lambda$ and $h_{\rm rms}$ parameters taken from 
 the AMF literature. In addition, we also obtain a suppression of the resistivity anisotropy (with respect to
 the case where strain is neglected) that consistent with Ref.~\onlinecite{Lundeberg10}.
We believe that the anisotropic nature of the random magnetic field disorder also significantly changes 
the quantum corrections to the conductivity with respect to the isotropic result \cite{Mathur01} and deserves 
further investigation. 

In summary, our results suggest that the investigation of anisotropy corrections to the Drude conductivity 
can be a new and insightful path to experimentally quantify effects of random pseudo-magnetic 
fields due to strain.

\acknowledgments 
We thank Eduardo Mucciolo for numerous discussions. This work has been 
supported by the Brazilian funding agencies CAPES, CNPq, and FAPERJ.

%


\end{document}